\UseRawInputEncoding 
\documentclass[prb,aps,twocolumn,groupedaddress,floats,showpacs]{revtex4}
\usepackage{graphicx}
\usepackage{dcolumn}
\usepackage{bm}
\usepackage{color}
\def\he4{$^4$He}
\def\hel3{$^3$He}
\def\Am3{\AA$^{-3}$}
\def\beq{\begin{equation}}
\def\eeq{\end{equation}}
\begin{document}
%%%%%%%%%%%%%%%%%%%%%%%%%%%%%%%%%%%%%% AUTHORS %%%%%%%%%%%%%%%%%%%%%%%%%
\author{Anatoly Kuklov$^1$, Emil Polturak$^2$, Nikolay Prokof'ev$^3$ and Boris Svistunov$^{3,4,5}$}
\affiliation{$^1$ Department of Physics and Astronomy, CSI, and the Graduate Center of CUNY, New York}
\affiliation{$^2$ Faculty of Physics, Technion-Israel Institute of Technology, Haifa, Israel}
\affiliation{$^3$ Department of Physics, University of Massachusetts, Amherst, MA 01003, USA}
\affiliation{$^4$ National Research Center Kurchatov Institute," 123182 Moscow, Russia}
\affiliation{$^5$ Wilczek Quantum Center, School of Physics and Astronomy and T. D. Lee Institute,
Shanghai Jiao Tong University, Shanghai 200240, China}
%%%%%%%%%%%%%%%%%%%%%%%%%%%%%%%%%%%%%%%%%%%%%%%%%%%%%%%%%%%%%%%%%%%%%%%%%%%%%%

\title{Thermal and structural properties of topological defects in solid \he4: Strain induced martensitic transformation from hcp to fc orthorhombic lattice  }

\date{\today}
%%%%%%%%%%%%%%%%%%%%%%%%%%%%%%%%%%%%%%%%%%%%%%%%%%%%%%%%%%%%%%%%%%%%%%%%%%%%%%
\begin{abstract}
Several experimental studies  have reported thermally activated behavior of the mechanical response of solid \he4 which does not fit into the model of thermally activated Frenkel pairs in an ideal crystal.  The purpose of the present work is to investigate how structural topological defects modify the response. Using quantum Monte Carlo Worm Algorithm, we study temperature dependence of fluctuations of the total number of particles in samples of hcp solid \he4 containing several types of dislocations.  Such fluctuations can be described by the thermal activation law with the activation energy dependent strongly on the type of the dislocation and the crystal density. The extracted values cover a range from about 1K to 20K. It is also found that annihilation of a jog-antijog pair can produce a local region of the solid characterized by the orthorhombic symmetry. This serendipitous observation suggests that
hcp solid \he4 can undergo a displacive phase transition into the orthorhombic crystal under uniaxial stress of about 10-15\%.
\end{abstract}

\maketitle

\section{Introduction}
Solid \he4 represents one of the most studied and still enigmatic strongly interacting many-body system. The ratio of a typical interaction and zero-point kinetic energies per particle is of the order of unity which excludes any controlled analytical treatment of the material.  This situation creates uncertainty for even qualitative predictions about the nature of its ground state. A strong early interest to solid \he4 has been inspired by the proposal that it is a supersolid \cite{Gross,Andreev,Thouless,Chester} -- that is, a phase combining properties of solid and superfluid. However, large scale experimental efforts, which, practically, immediately followed this proposal, yielded no supportive evidence. The interest to the subject has reemerged after the experiments  \cite{KC} have found a possible evidence of the supersolid phase in the temperature variation of the torsional oscillator frequency. This variation, however, was later shown to be unrelated to the supersolidity and was explained in terms of the dislocation dynamics \cite{Beamish_Nature}. 

In the absence of reliable analytical tools for studying solid \he4, {\it ab initio} numerical methods based on the path integral formulation of quantum mechanics  
became especially important (see in Ref. \cite{Cep1995}). Immediately after the observation of the torsional oscillator anomaly \cite{KC}, several groups have unambiguously established that the ideal solid \he4 is not a supersolid \cite{Cep,WA,Clark}. Using a generic argument \cite{PS2005} it has been shown that zero-point vacancies or interstitials must be an integral part of the ground state in order for a solid to become supersolid. Thus, in the absence of the symmetry between vacancies and interstitials, a commensurate solid in free space has measure zero to be a supersolid. Finally, direct simulations of several vacancies introduced into an ideal solid \he4 sample have found that they phase separate into hexatic loops at low temperatures \cite{fate}, and, thus, cannot form the superfluid condensate envisioned in Ref.\cite{Andreev}.  

Despite that no predicted supersolid phase has been found, it is important to realize that strong quantum fluctuations in solid \he4 did bring surprising results not found in any other known material.
Structural defects -- some grain boundaries \cite{GB} and screw dislocation (with Burgers vector along the C-axis) \cite{screw} -- have been found in the {\it ab initio} simulations to support (quasi) 2D and 1D superfluidity, respectively.   
A couple of years later a flow of \he4 atoms through solid \he4 has been detected in the experiment conducted by the UMass group \cite{Hallock}. Typical characteristics of this flow -- temperature and bias dependencies -- are clearly inconsistent with any type of classical dissipative dynamics. The flow rate shows subohmic dependence on the bias and it decreases as temperature increases. These features have been confirmed by other groups \cite{Beamish,Moses}. 
It is important to note that the flow observed in Refs.\cite{Hallock, Beamish} is accompanied by another unexpected feature -- a so called {\it syringe} effect, when matter accumulates inside the solid biased by the chemical potential. This feature turned out to be consistent with the so called {\it superclimb} of edge dislocations (with their Burgers vector oriented along the C-axis) observed in the {\it ab initio} simulations \cite{sclimb}.   

Current understanding of the superflow-through-solid \cite{Hallock,Moses} and the syringe \cite{Hallock,Beamish} effects is based on the assumption that the solid \he4 contains a network of dislocations with superfluid cores \cite{shevchenko} providing pathways for the superflow. Such a network would be unstable due to the superclimb effect, but other (non-superfluid) dislocations provide a stabilizing framework \cite{network} -- due to the binding between the superfluid and non-superfluid edge dislocations \cite{Kuklov2019}. 
At this point it is important to emphasize that the hcp solid \he4 displays also an unusual mechanical response  characterized by so called giant plasticity \cite{giant_plast} as well as strongly non-linear behavior \cite{non_linear,RMP} of basal dislocations.

 It has also been found that at temperatures high enough to prevent binding of
\hel3 impurities to dislocations,   the mechanical response can be interpreted in terms of activation energies well below typical energies for creation of vacancies or interstitials \cite{Beamish_5,Polturak_3_6}. More recently yet another value of the activation energy significantly below the vacancy creation energy has been reported for the so called "anomalous samples" in Ref.\cite{Iwasa}. These observations call for more quantitative studies of the non-superfluid dislocations. 

Here we focus on thermal properties of edge dislocations rather than on the superfluidity along their cores.
These include basal dislocations
(that basal dislocations don't have superfluid cores has been reported in Refs.\cite{screw,GB,nonSF,Borda}), as well as the non-basal ones which have Burgers vector along the basal plane and the core aligned with the the hcp axis. [The elementary segment of such a dislocation represents a jog on the basal dislocation]. The partial dislocation, which is a boundary of the structural E-fault (see in Refs.\cite{Hirth}) and is characterized by 1/2 of the Burgers vector along the hcp axis, has been studied as well. At $T\leq 0.5$K and densities close to melting $\sim 0.0287 \AA^{-3}$, this partial has been found to have a superfluid core \cite{sclimb}. At larger density, $\sim 0.03 \AA^{-3}$, and temperatures $T\geq 0.5$K the core superfluidity vanishes and is replaced by thermally activated response.

Our main findings are as follows. There is a wide spectrum of the thermally activated response characterized by activation energies $E_a$ covering the range from 1K to 20K. The values of $E_a$ strongly depend on the dislocation type and crystal density -- the density increase by only about few percents results in the increase of the activation energy $E_a$ by a factor of 2.  %Basal jogs carry a thermal cloud (of a radius of about 10-15\AA) of the point-like defects characterized by the activation energy $ E_a=(5.2 \pm 0.8)$ K at density $0.0288\AA^{-3}$ . Basal (jogless) dislocations have activation energy $E_a=(7.5 \pm 1.5)$ K  at the density $0.0288\AA^{-3}$, and $E_a$  increases by a factor of 2 once the density is raised by only about 4\%. The E-fault partial is characterized by  $E_a =(2.5 \pm 1)$K. These all values of $E_a$ should be compared with $E_a =(19\pm 1)$K observed for the ideal hcp solid \he4 at low density, $0.0288 \AA^{-3}$.  
Specific values of $E_a$  are summarized in Sec.\ref{fits}. 

Simulations of the basal dislocations and their jogs have been conducted with periodic boundary conditions on samples containing about 2000 particles. Accordingly, the topological defects were present in pairs of opposite "charge". This opened up a possibility for the annihilation of the dislocation pairs. Despite this, no annihilation has been detected for the pair of basal dislocations. In contrast, the non-basal pairs (representing a jog-antijog pair on the basal dislocation) did annihilate on long runs at temperatures above 1K. As a result, the sample either acquired additional particles and became a slightly stretched hcp solid or lost particles and transformed into a solid of orthorhombic symmetry. We suggest the following mechanism for this transformation: the lost half plane of particles results in stretching of the hcp crystal along its (1,0,0) direction, and this strain induces the phase transition changing the symmetry.  Given the size of the simulated samples the estimated critical strain turns out to be about $10-15\%$.     This observation motivates further systematic studies of the solid \he4 structure under strain.

\section{Theoretical framework}
At finite temperature a crystal acquires a finite number of vacancy-interstitial pairs -- Frenkel defects. These defects are interacting with dislocations and control their climb \cite{Hirth} (as opposed to glide which can occur without the point-like defects). This results in disentangling dislocation lines and can eventually destabilize any dislocation network. Accordingly, our focus here is on measuring thermal response of the solid with dislocations on inserting or removing particles.
 In an ideal crystal a number $N_p$ of Frenkel defects at low temperature $T$ is well approximated by the thermal activation dependence 
\beq
N_p =N_0 {\rm e}^{-\frac{E_v + E_i}{T}},
\label{activ}
\eeq
 where $N_0$ stands for the total number of atoms in a crystal and $E_v,\, E_i$ denote energies for creating a vacancy and interstitial, respectively, relative to the chemical potential of the system in the Grand Canonical ensemble (GCE).  This dependence is typical for classical particles thermally excited from their ideal crystal lattice positions. Point defects in solid Helium are quantum objects. However, in the main exponential approximation their activated fraction can also be evaluated using Eq.(\ref{activ}). These energies have been calculated in Ref.\cite{fate} as $E_v =(13.0 \pm 0.5)$ K, $ E_i =(22.8 \pm 0.7)$ K in the ideal hcp crystal at the melting pressure.
This result is based on evaluating Matsubara single-particle Green's function at low temperature (0.2K).

Here we have  adopted a different strategy. 
In an ideal crystal at equilibrium values of the chemical potential $\mu$ the total mean  number of particles $\langle N \rangle $  changes as a function of $T$ according to the relation  $\langle N \rangle = N_0 - N_v + N_i $, where $N_{v,i}\approx N_0 \exp(-E_{v,i}/T)$ are the mean  numbers of vacancies and interstitials, respectively, and $N_0$ stands for $\langle N\rangle $ at $T=0$. Thus, if a system is characterized by $E_v \neq E_i$, which is a consequence of  lack of a particle-hole symmetry, it would be possible to measure the point defect number characterized by the smallest activation energy  in the main exponential approximation, provided $E_i$ and $E_v$ are not too close to each other. As mentioned above, in the ideal solid \he4 the activation energy of interstitials is significantly larger than that of the vacancies. Thus, this relation simplifies as  $\langle N \rangle = N_0 - N_0\exp(-E_v/T) $, and measurements  of $\langle N \rangle$ versus $T$ can provide $E_v$. The relation between $E_i$ and $E_v$ close to a dislocation is not known apriori. Furthermore, the very notions of vacancies and interstitials become poorly defined because of the processes of particles being added  (or subtracted) to (from) the dislocation core which results in the dislocation climb \cite{Hirth}. In particular, it is clear that such processes must be characterized by the same activation energy -- by the token of the symmetry with respect to the direction of the climb.   Hence, to avoid the cancellation of $N_v$ and $N_i$ in this case, we have focused on the mean squared fluctuations of $N$ which take into account both types of defects on equal footing. Specifically, the quantity   $\sigma^2_N = \langle N^2 \rangle  - \langle N\rangle^2$ has been evaluated for different temperatures in GCE. These fluctuations are determined by the total numbers of the thermally excited defects in the system. At low density (at temperatures much higher than a possible degeneracy temperature) these defects behave as an ideal gas. Thus, one can use the relation 
\beq
\sigma^2_N=N_v + N_i.
\label{NvNi}
\eeq            
 As will be discussed below, $\sigma^2_N$ does show the activated dependence 
\beq
\sigma_N^2 \propto {\rm e}^{-\frac{E_a }{T}},
\label{sigmaE}
\eeq
with $E_a$ covering a wide range -- from 1K to 20K. It is important to realize that such point defects with the activation energy below the value typical for the ideal crystal are bound to dislocations and form a cloud  of normal fluid surrounding dislocation cores.  

It should also be emphasized that not all the dislocations can be characterized by the thermally activated behavior at low $T$.  Dislocations showing the superfluid response along their core do not fit into the above framework. In particular, the $E$-fault dislocation shows the two stage activation. This feature will be discussed in more detail later.

\section{Samples and simulations}
Initial samples  have been prepared starting from atomic positions arranged accoring to the ideal hcp symmetry. In order to produce a topological defect, a corresponding 1/2 plane of atoms has been removed and the remaining spatial gap healed by means of purely classical simulations with some repulsive interaction potential between atoms. Then, simulations have been conducted by the Worm Algorithm \cite{WA} for several temperatures starting from T=0.25 K and up to 2.5K at corresponding values of the chemical potential $\mu$ and two different  densities $0.0288$\AA$^{-3}$ and $0.030$\AA$^{-3}$. %corresponding to pressures 25 bar and 39 bar, respectively. 
Quantum configurations of the world lines of the atoms in imaginary time $\tau \in (0,1/T)$ were periodically projected into 
classical positions. As a qualitative assessment of the role of quantum fluctuations, the map of particle exchanges has been superimposed atop of the classical snapshot of the positions. This map (shown by blue stars in figures below) has served as an imaging tool for the most probable areas where particles were introduced or removed.

\subsection{Ideal hcp}
\begin{figure}[!htb]
%\vskip-8mm
	\includegraphics[width=1.1 \columnwidth]{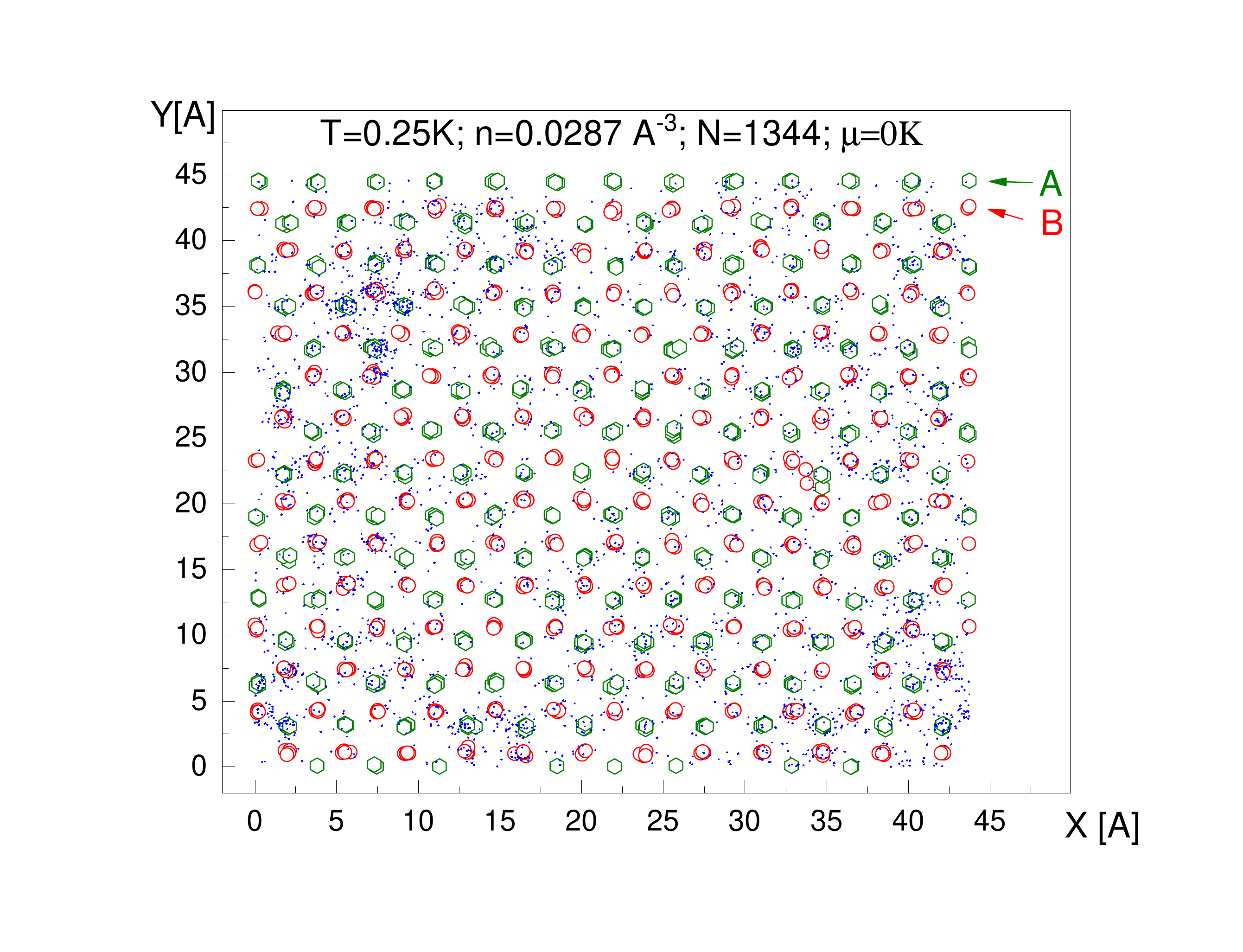}
	\vskip-8mm
	\caption{(Color online) One (typical) snapshot of atomic positions in hcp solid as viewed along the high symmetry axis (Z-direction). The A,B triangular layers are shown by (green) open hexagons and (red) open circles, respectively. %Deviations from the ideal positions of triangular layers are due to quantum and thermal fluctuations. 
The (blue) stars indicate places where quantum exchanges take place.}
	\label{fig1}
\end{figure}

The hcp structure is formed by two identical triangular layers A and B shifted with respect to each other and stack together along the hcp axis in the ABABAB... order (see in Refs.\cite{Hirth}). A columnar view along the hexagonal axis of such a crystal is shown in Fig.\ref{fig1}.
 The projection of the atomic positions on one plane normal to the hcp axis produces a 2D image of a honeycomb lattice, where (green) hexagons represent A-layers and (red) circles -- B-layers. 
Visible deviations of atoms from equilibrium positions are due to large quantum fluctuations in solid \he4. For the density $0.0288 \pm 0.0001\AA^{-3}$ the GCE particle number fluctuations suggest that $E_a =(19\pm 1)$K,  see Fig.~\ref{fig10}.

\subsection{Pair of basal edge dislocations}
A pair of basal edge dislocations with opposite topological charges has been created by cutting out one half of full plane of atoms perpendicular to the direction [1,0,0] (X-direction in Fig.\ref{fig1}) in the hcp ideal crystal with the periodic boundary conditions (PBC) in all directions. 
\begin{figure}[!htb]
%\vskip-8mm
	\includegraphics[width=1.1 \columnwidth]{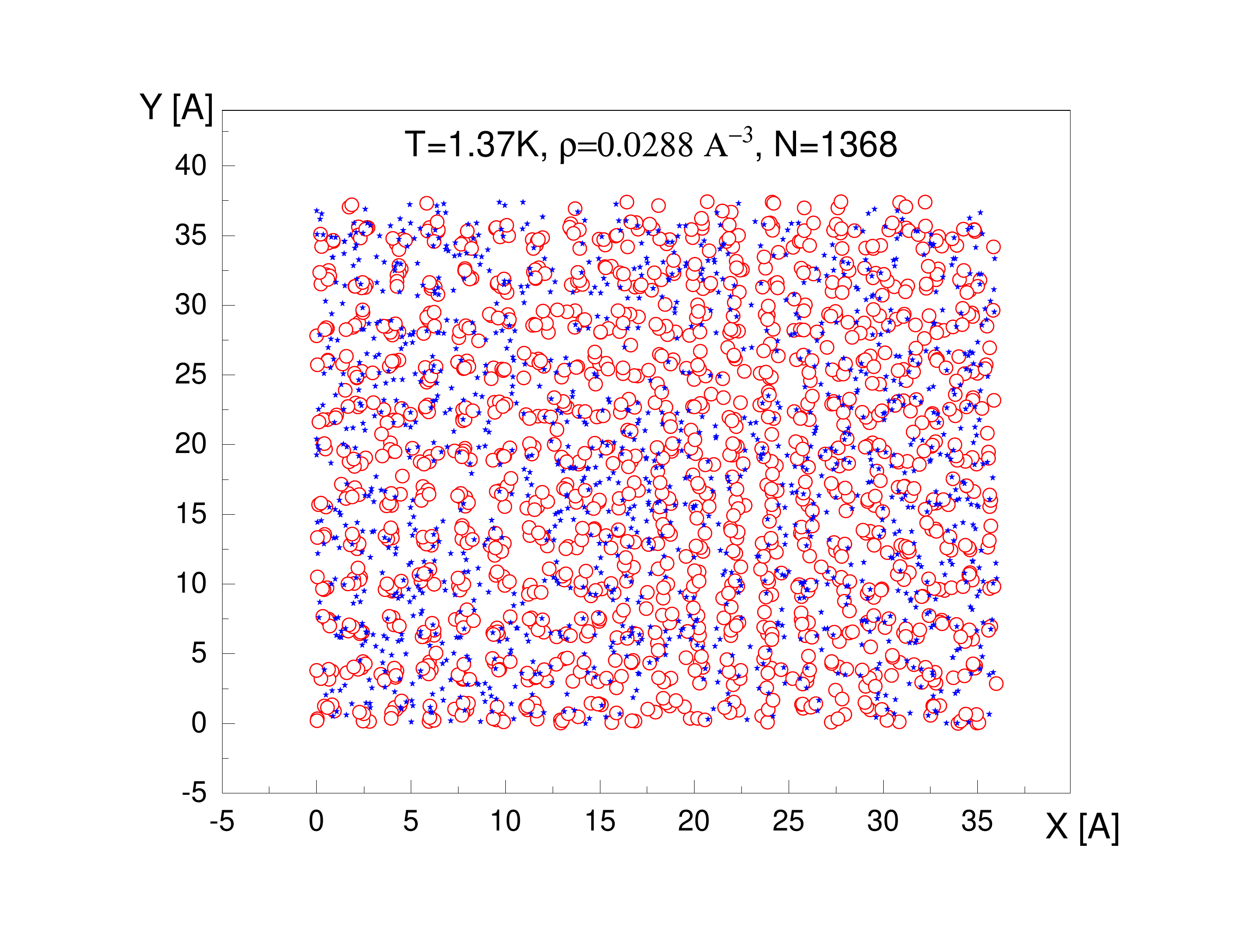}
	\vskip-8mm
	\caption{(Color online) One (typical) snapshot of atomic positions (open red circles) forming hcp solid containing a pair of basal edge dislocations as viewed along the high symmetry axis [0,0,1] (Z-direction). }
	\label{fig3}
\end{figure}
\begin{figure}[!htb]
%\vskip-8mm
	\includegraphics[width=1.1 \columnwidth]{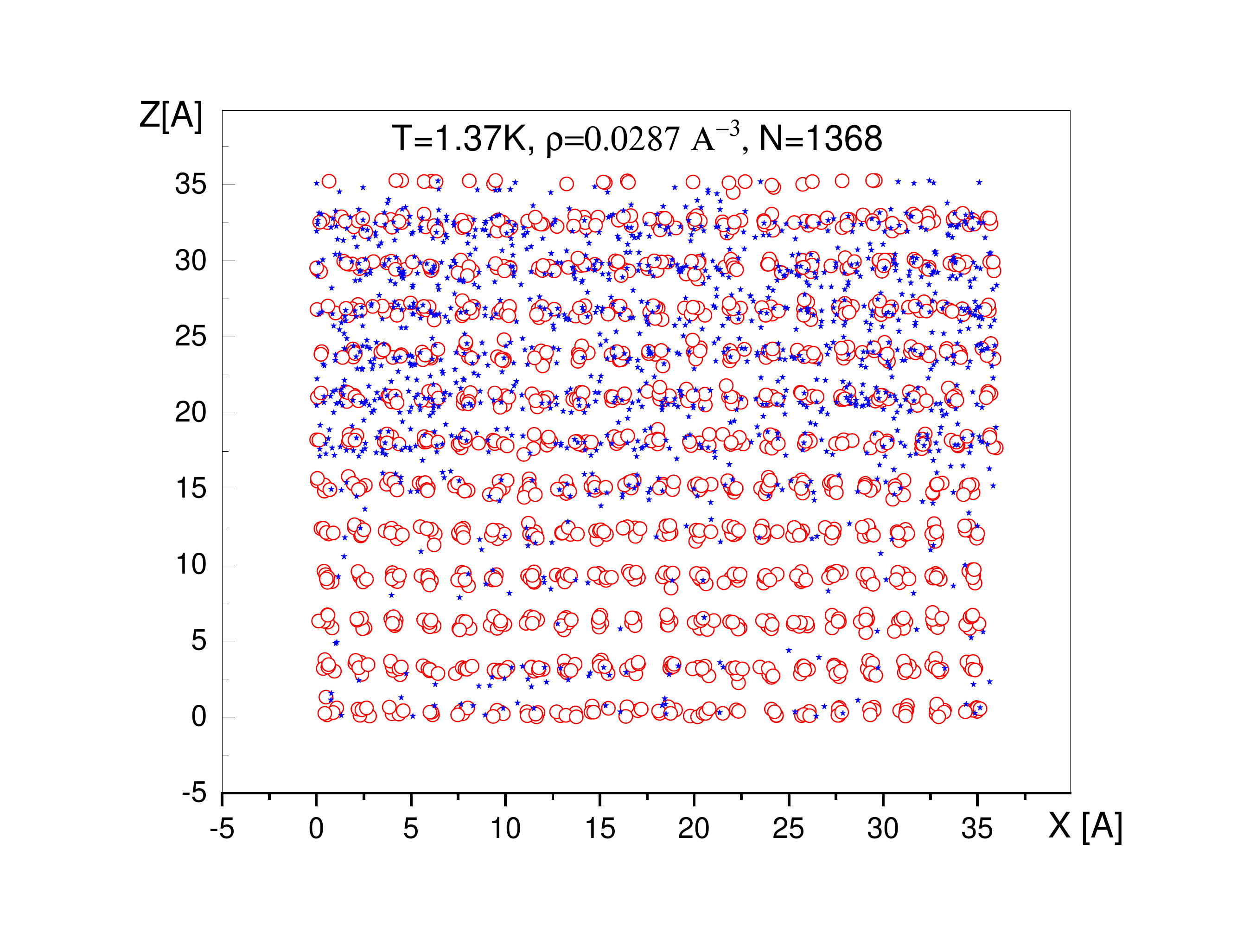}
	\vskip-8mm
	\caption{(Color online) The columnar view along the [0,1,0] direction (Y-axis) of the same sample as in Fig.\ref{fig3}.}
	\label{fig4}
\end{figure}
\begin{figure}[!htb]
%\vskip-8mm
	\includegraphics[width=1.1 \columnwidth]{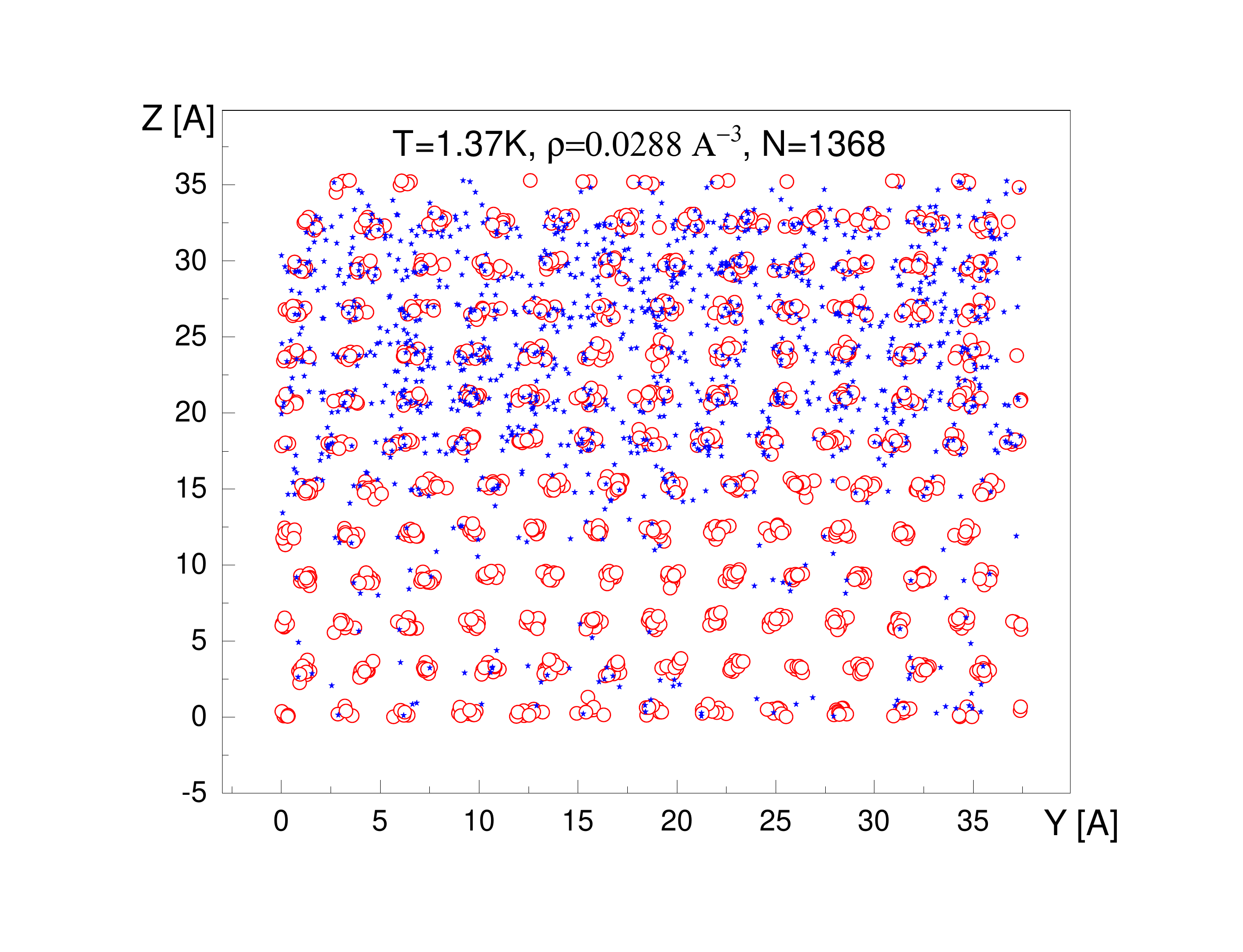}
	\vskip-8mm
	\caption{(Color online) The columnar view along the [1,0,0] direction (X-axis ) of the same sample as in Fig.\ref{fig3},\ref{fig4}.}
	\label{fig5}
\end{figure}
The edges of the remaining 1/2- plane are full basal dislocations with opposite charges. Because of the PBC these dislocations can be viewed as two loops. Such an initial configuration has been used as an input for the WA algorithm. In the course of the simulations each full basal dislocation split into two partials (cf. \cite{Borda}).
One snapshot of the resulting configurations is shown in Fig.~\ref{fig3},\ref{fig4},\ref{fig5} as viewed along 3 perpendicular directions. As can be seen, there is little of the (columnar) honeycomb order left, Fig.~\ref{fig3}, as compared with the ideal solid, Fig.~\ref{fig1}. We attribute this to splitting of the dislocations into partials and their large shape and positional fluctuations along the basal plane. It is important to note that the crystalline order is well preserved along other directions  (see Figs.~\ref{fig4}, \ref{fig5}). 

At this point we note that the exchange map shown in Figs.~\ref{fig1}, \ref{fig3}, \ref{fig4}, \ref{fig5} as (blue) stars  is a statistical quantity -- that is, it is collected over many quantum configurations. In contrast, the positional snapshot reflects a particular atomic arrangement (which is the $\tau$-averaged configurations of the atomic worldlines). As seen in Figs.~\ref{fig4},\ref{fig5}, density of the exchanges is higher in the part of the sample where the half-plane of atoms has been eliminated, and, thus, the density of atoms is lower (at $Z \geq 15\AA$). It is worth mentioning the uniformity of the map along X,Y directions. If the dislocations (4 partials) were fixed in one place, the map would be located mostly in the vicinity of their cores. [This effect will be discussed later for other dislocations]. The uniformity of the map indicates that the dislocation partials sweep the sample along the XY planes many times  during the period when the map was collected. 
The activation energies obtained from these simulations are $E_a=7.5 \pm 1.5$ K and $E_a=15\pm 1$K for the crystal densities $\rho=0.0288\AA^{-3}$ and $\rho=0.0300 \AA^{-3}$, respectively;  i.e., the activation energy doubles by increasing density by only $4 \%$. 

\subsection{Jog-antijog pair }
\begin{figure}[!htb]
%\vskip-8mm
	\includegraphics[width=1.1 \columnwidth]{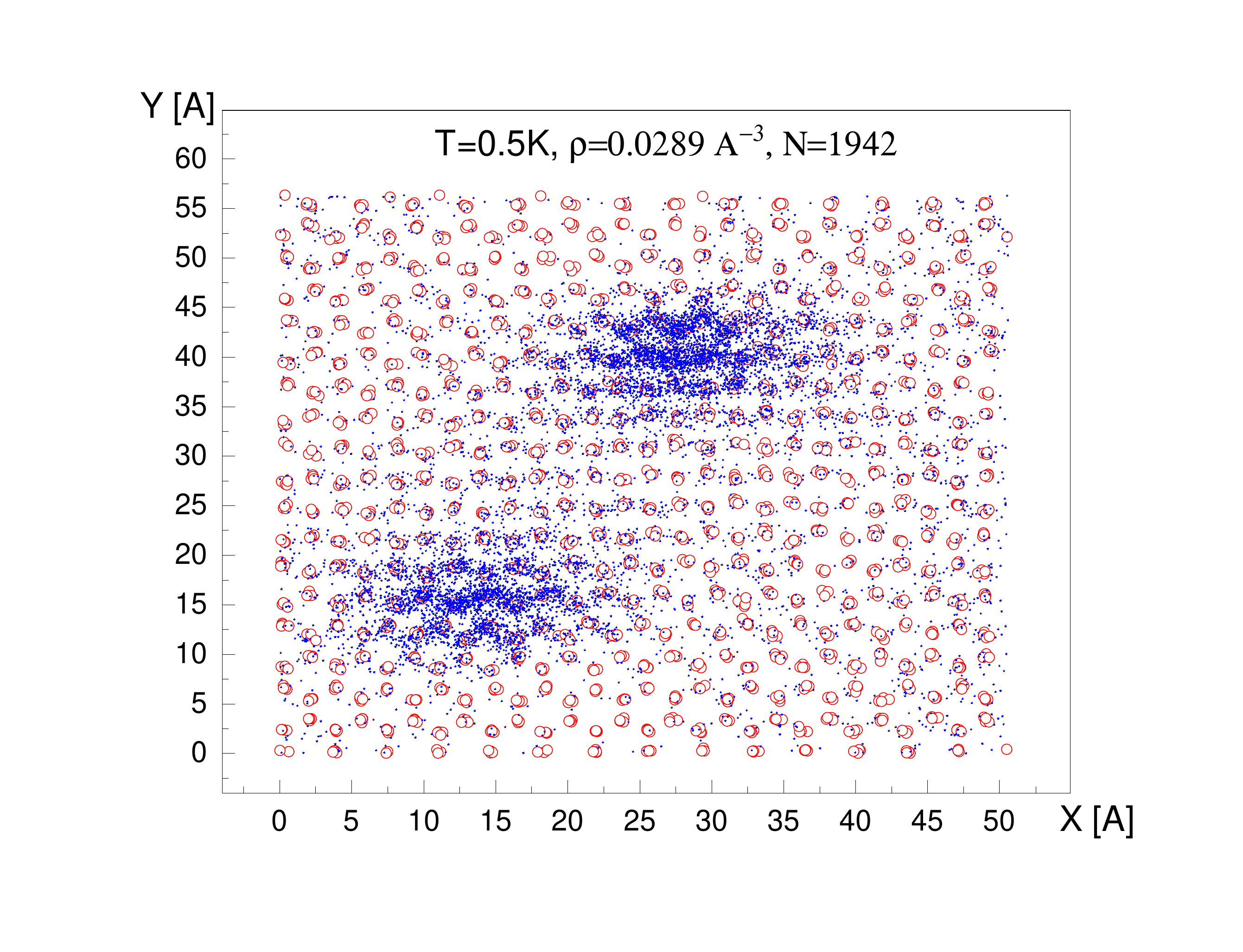}
	\vskip-8mm
	\caption{(Color online) One (typical) snapshot of atomic positions (open red circles) in a sample containing a pair of non-basal edge dislocations 
with their cores oriented along the Z-axis (and Burgers vector in the basal plane) and located within the high density (blue) clouds of the exchange map.}
	\label{fig6}
\end{figure}

\begin{figure}[!htb]
%\vskip-8mm
	\includegraphics[width=1.1 \columnwidth]{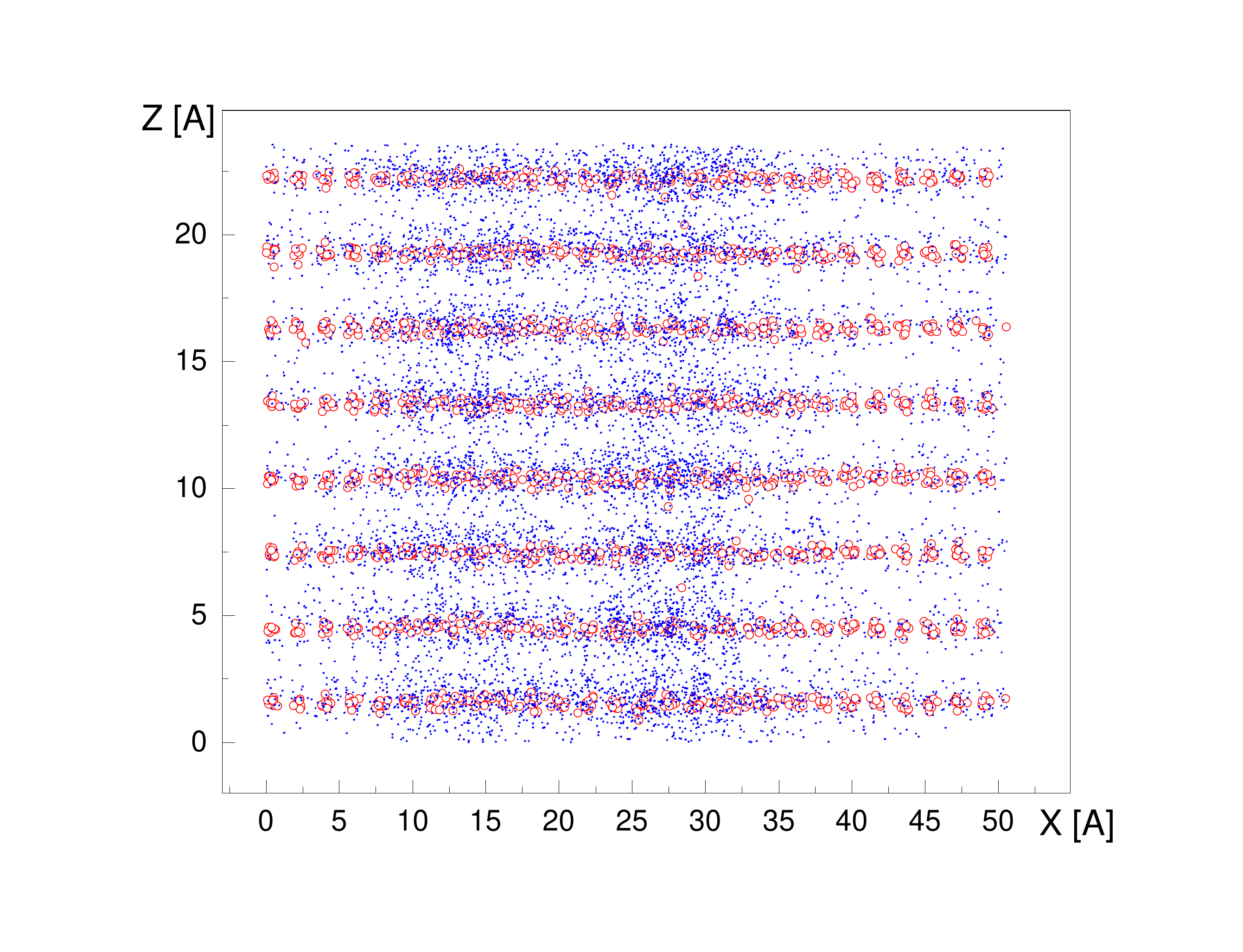}
	\vskip-8mm
	\caption{(Color online) The columnar view along the [0,1,0] direction (Y-axis) of the same sample as in Fig.\ref{fig6}.}
	\label{fig7}
\end{figure}

\begin{figure}[!htb]
%\vskip-8mm
	\includegraphics[width=1.1 \columnwidth]{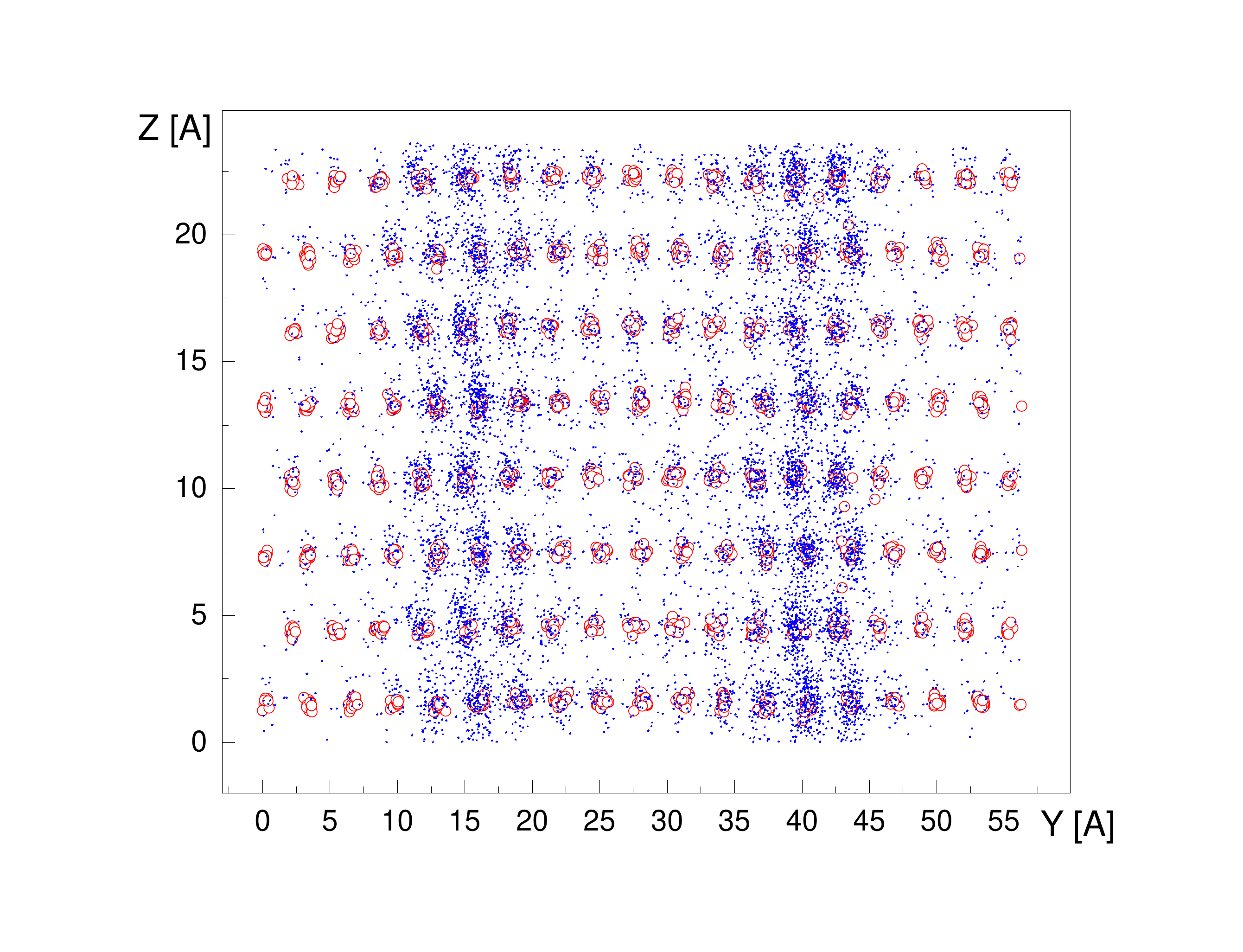}
	\vskip-8mm
	\caption{(Color online) The columnar view along the [1,0,0] direction (X-axis ) of the same sample as in Fig.\ref{fig6}.}
	\label{fig8}
\end{figure}

A basal edge dislocation can have kinks and jogs. A kink corresponds to an elementary deformation of the dislocation line  within the basal plane. A jog is created when two parts of a basal dislocation (or its partial) belong to neighboring basal planes. Thus, an elementary jog can be viewed as a part of the core of the edge dislocation oriented along the Z-axis while its Burgers is along the basal plane. A typical snapshot of the sample containing a pair of such dislocations is shown in Figs.~\ref{fig6},\ref{fig7},\ref{fig8} from different projections.
\begin{figure}[!htb]
%\vskip-8mm
	\includegraphics[width=1.0 \columnwidth]{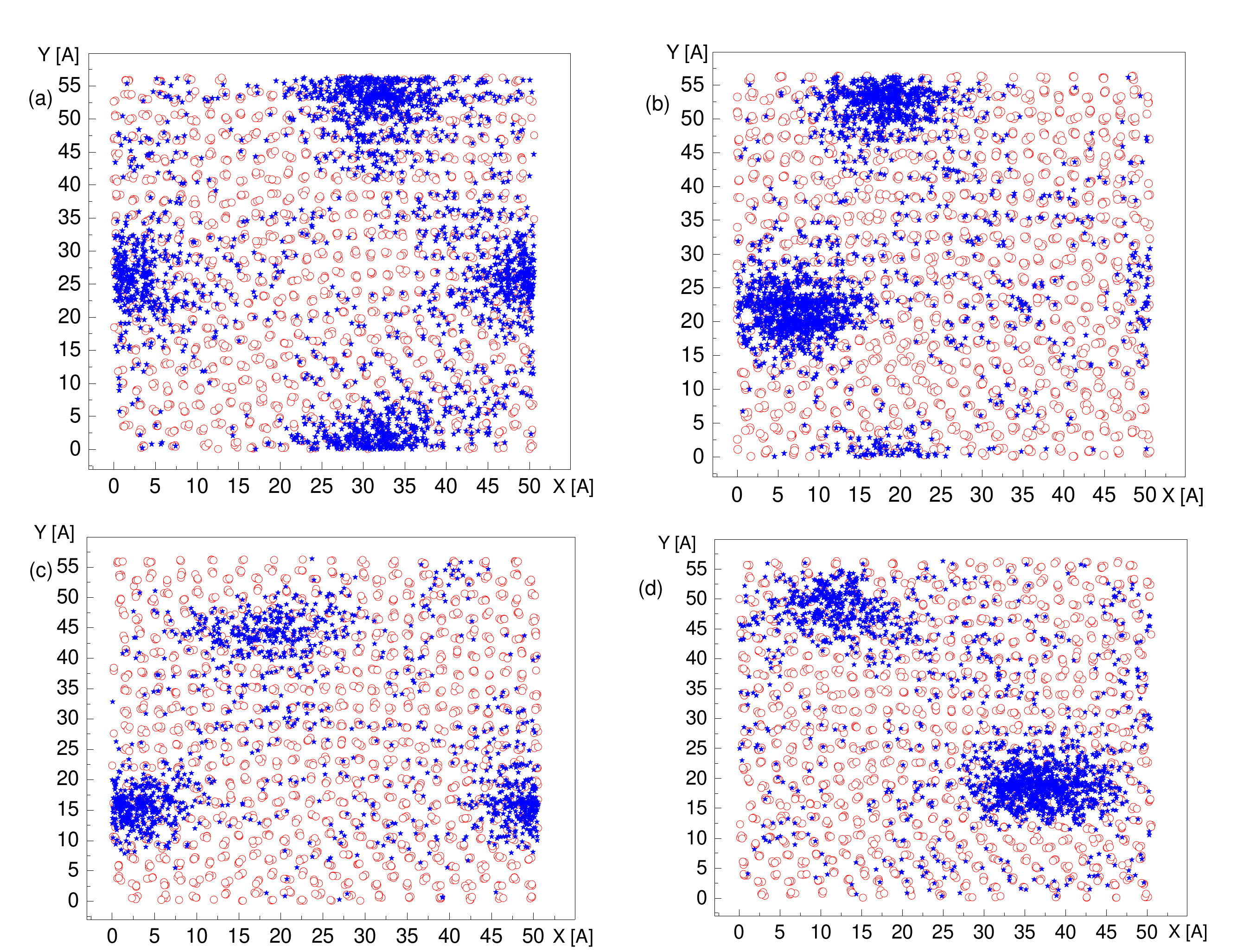}
%	\vskip-8mm
	\caption{(Color online) Snapshots of the same sample at $T=1.0$K taken with the intervals of 5$\cdot 10^7$ MC updates per particle in the progression of (the Monte Carlo) time from (a) to (b) to (c) to (d). The exchanges map was collected for 5$\cdot 10^6$ updates per particle.  The clouds of (blue) stars mark positions of the jog and antijog. Once a cloud reaches a boundary, it splits into two due to the periodic boundary conditions. }
	\label{fig9}
\end{figure}

In contrast to the basal dislocation pair, the jogs are not that mobile. Their "motion" as a result of the MC updates is shown in Fig.\ref{fig9}. It takes about $10^7-10^8$ updates per particle in order to see a visible shift of the pair position. Accordingly, the exchanges map is concentrated at their positions and serves as the imaging tool.  Since there are only two clouds, it can be concluded that the full dislocation does not split into partials (or its splitting is under 10 \AA).  [Four and three clouds seen in Fig.\ref{fig9} in panels a,b,c is a result of the periodic boundary conditions]. The activation energy in this sample was found to be $E_a=(5.2 \pm 0.8)$ K, Fig.\ref{fig10}.

We note that the jog pair can annihilate -- in about $10^9$ updates per particle. Depending on the value of chemical potential, the sample either gains some number of particles (to heal the missing half-plane of atoms) or looses particles and leaves the sample stretched by about 10-15\%. It turns out that such a  deformation results in changing the sample symmetry -- from hcp to the face centered orthorhombic. We will discuss this symmetry change later.   

\subsection{Basal fault edge dislocation}
\begin{figure}[!htb]
%\center
\vskip-8mm
	\includegraphics[width=1.1 \columnwidth]{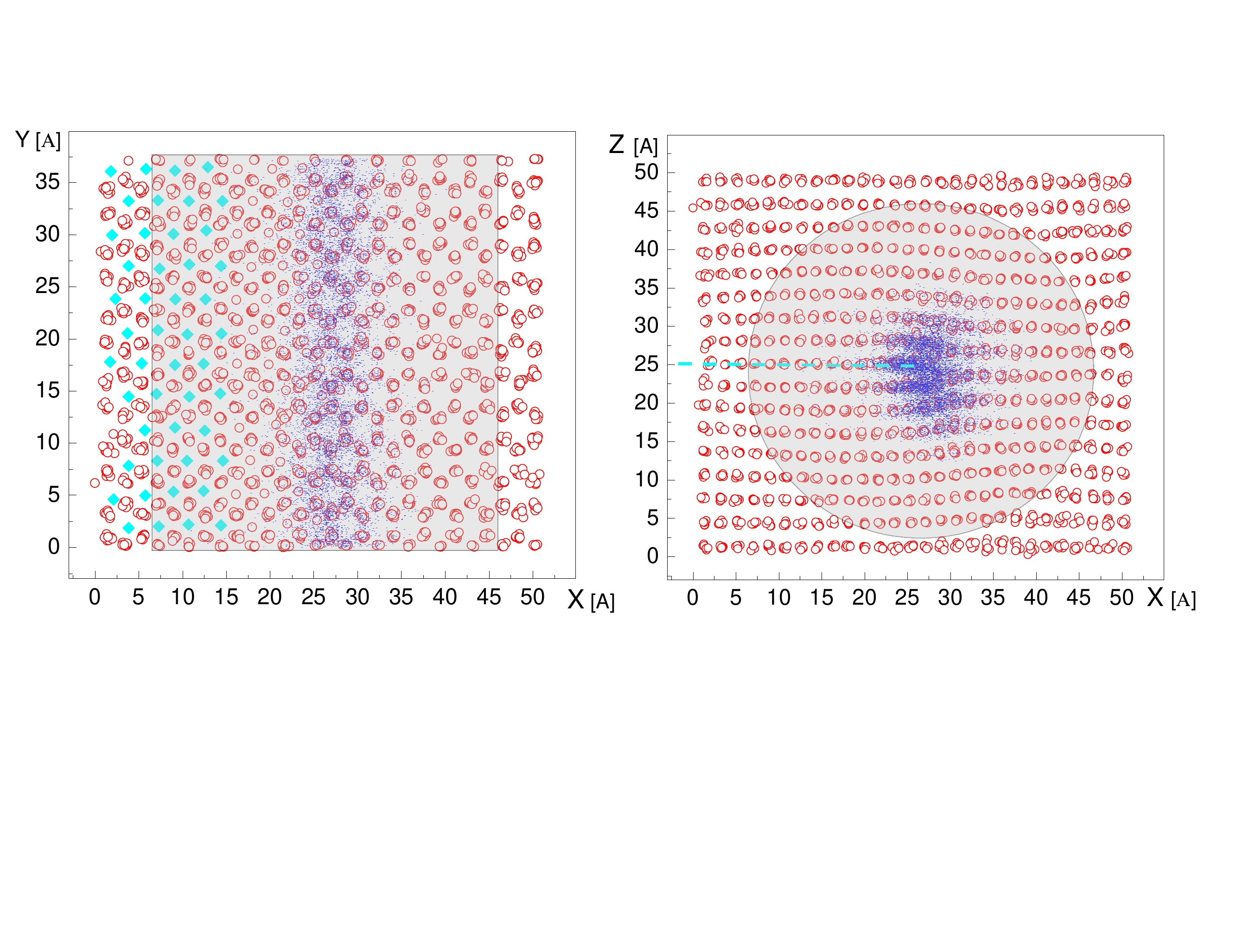}
	\vskip-12mm
	\caption{(Color online) Atomic positions (open circles) of a sample containing the E-fault edge dislocation with the exchange map. The dimmed areas indicate the cylinder containing fully updatable particles. Particles outside this cylinder are frozen. Left panel: Columnar view along the hcp axis. The fault incomplete plane is partially shown by (cyan) squares. Right panel: Columnar view along the dislocation core (along the Y-axis). The E-fault incomplete plane is marked by the (cyan) dashed line.  }
	\label{fig11a}
\end{figure}
The E-fault in the hcp structure is formed when in the  ABABABAB... stack of atomic layers there is one inserted triangular layer C, say, ABABCABAB... \cite{Hirth}. This ABC element of 3 layers has the fcc symmetry. 
 If the plane C is incomplete, the edge of the C-plane represents one partial of the full edge dislocation with the Burgers vector along the hcp symmetry axis. This full dislocation has a superfluid core at low temperature -- as observed in the simulations \cite{sclimb}. Furthermore, it splits into two partials one of which is the edge of the E-type fault. Here we study the thermal properties of this partial at density $0.03 \AA^{-3}$ -- when there is no significant superfluid response along the core.   A typical snapshot of the atomic positions of such a dislocation is shown in Fig.\ref{fig11a}.

At this point we would like to comment on the sample preparation procedure. In order to prevent the annihilation of the dislocation pair with the opposite "charges", full quantum updates have been applied to only one dislocation -- inside the cylinder shown in Fig.~\ref{fig11a}. This was achieved by freezing out atoms (together with the "antidislocation") outside the cylindrical region. These  particles have been annealed first as though if they were distinguishable (by excluding updates with quantum exchanges) and, then, have their worldlines frozen for the rest of the simulation.  As a result, the periodic boundary conditions are satisfied only along the dislocation core inside the cylinder (the Y-axis in Fig.\ref{fig11a}). 

As has been mentioned above, the thermal response of this dislocation is different from other dislocations. It shows the two stage activation characterized by  $E_a \approx 1$K  in the range $0.25$K$\leq T\leq 1.37$K, and, then, $E_a \approx 4$K for $T>1.37$K.    It is important to note that the  data point at $T=0.25$K shows $\sigma_N^2$ a factor of 3 larger than it would be expected from the simplistic model of classical activation. [This data point is surrounded by the dimmed circle in Fig.~\ref{fig10}]. In addition, there is a week superfluid response (due to the finite length of the core). This features are direct consequences of the proximity to the superfluid phase of the core.  We interpret the first stage as a crossover between the quantum and classical fluctuation regimes.

\subsection{Temperature dependencies}\label{fits}
\begin{figure}[!htb]
%\vskip-8mm
	\includegraphics[width=1.0 \columnwidth]{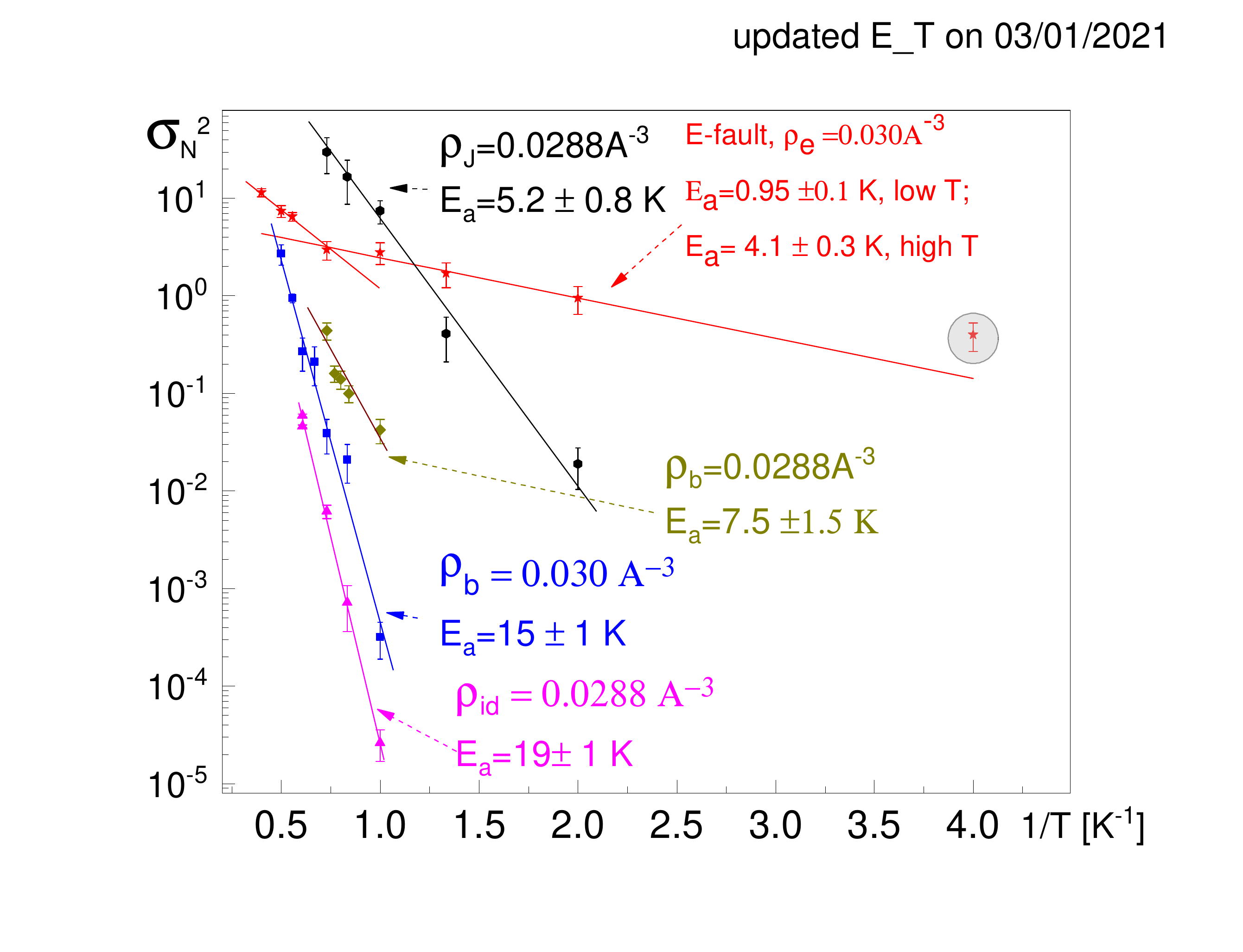} %Ea_Ybasal_den0299.pdf}
%	\vskip-8mm
	\caption{(Color online) Particle number fluctuations $\sigma_N^2$ determined for: a pair of basal dislocations at two densities $\rho_b =0.0288 \AA^{-3}$ (dark yellow rhombi), $ 0.030 \AA^{-3}$ (blue squares) ; an ideal hcp solid of density $\rho_{id} = 0.0288 \AA^{-3}$ ( pink triangles); the E-fault partial at density $\rho_e =0.030 \AA$ (red stars); a pair of non-basal edge dislocations with their cores along the Z-axis ( black hexagons). The corresponding activation energies and the densities are mentioned next to each line. The data point inside the dimmed circle is off the activation dependence by about a factor of 3.}
	\label{fig10}
\end{figure}

Temperature dependencies of $\sigma_N^2$ for the simulated samples are presented in Fig.\ref{fig10}. 
It is interesting to note that the activation energy $E_a$ in the presence of the basal dislocation in a sample of lowest density is a factor of 3 lower than in the ideal hcp solid. Furthermore, $E_a$ depends strongly on the sample density. Increase of the density by about 4\% results in the  increase of the activation energy by a factor of 2. 

The thermal behavior of the E-partial is characterized by significantly lower value of $E_a$ at $T\leq 1.37$K.  This can be interpreted in terms of the mechanism leading to the superfluidity of the dislocation core at low $T$ and low densities \cite{sclimb} -- due to the closing of the vacancy gap as induced by the local strain around the core \cite{strain}. Such a critical strain has been estimated to be about 10-12 \% at the lowest solid densities. At higher density such a strain may become insufficient  and the gap remains finite. Thus,  this low $E_a$ should reflect the value of the remaining energy gap for vacancy creation which can be estimated as $E_a \approx 1$K.

\section{Structural transformation induced by the jog pair annihilation}
\begin{figure}[!htb]
\vskip-8mm
	\includegraphics[width=1.0 \columnwidth]{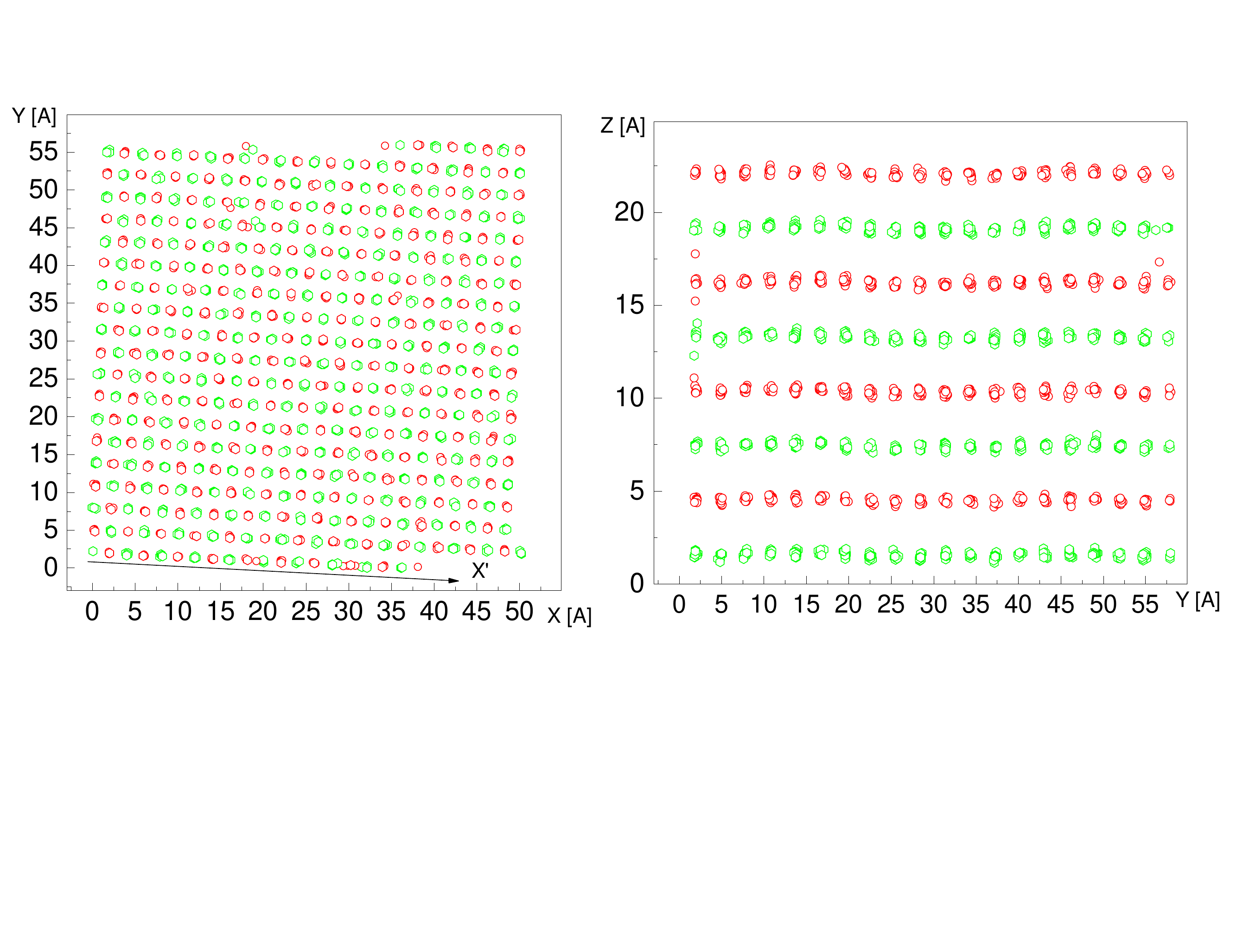}
	\vskip-15mm
	\caption{(Color online) The crystal structure resulting from the annihilation of the jog-antijog pair. Left panel: The columnar view along the original hcp-axis. Open (green) hexagons mark A-layers and open (red) circles -- B-layers. Right panel: The view along the X'-axis which is rotated by about -3$^o$ with respect to the original X-axis. }
	\label{fig11}
\end{figure}
\begin{figure}[!htb]
\vskip-8mm
	\includegraphics[width=1.2 \columnwidth]{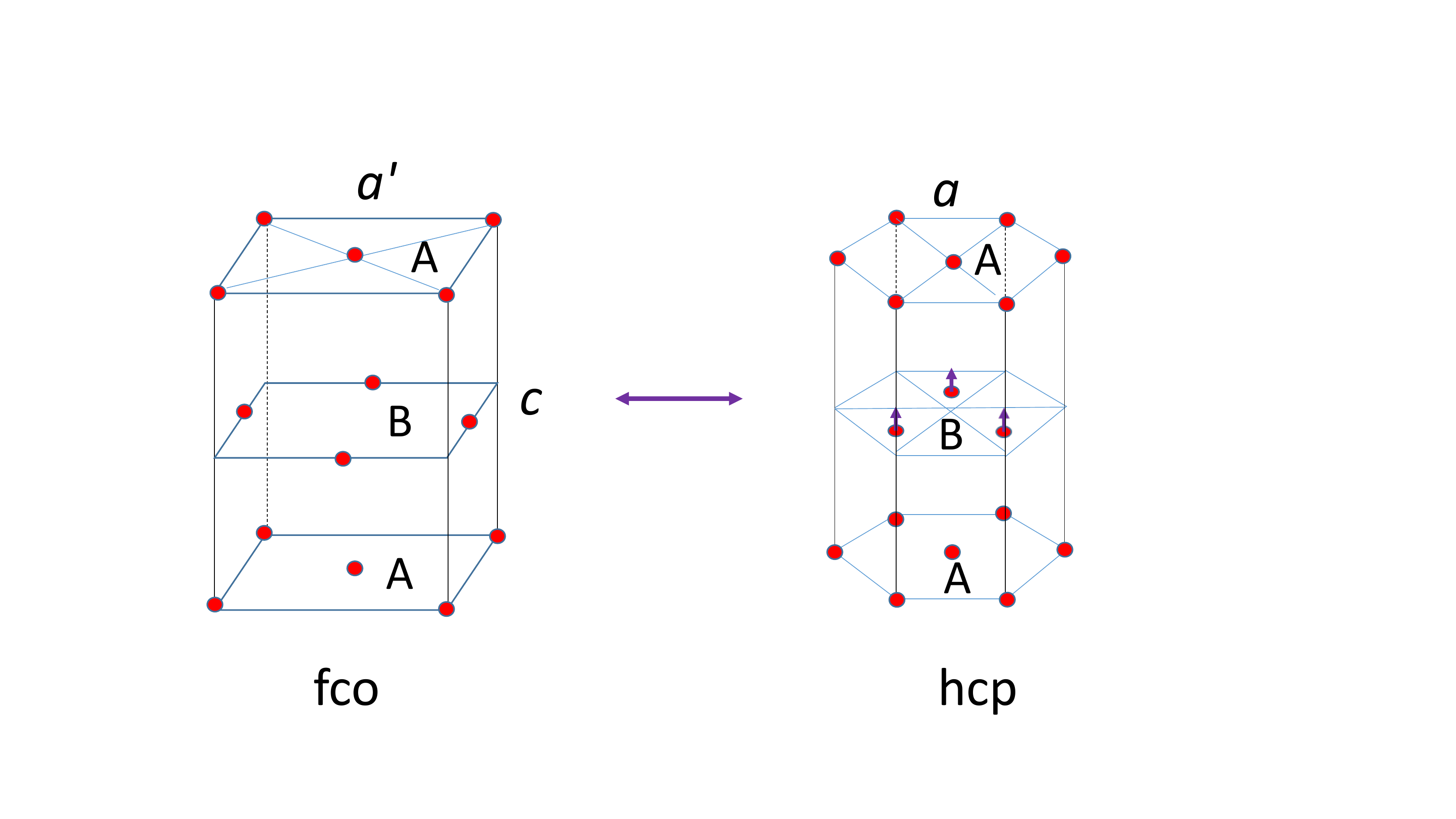}
%	\vskip-15mm
	\caption{(Color online) The unit cells of the hcp and the fco crystals. The fco structure is obtained from the hcp by shifting the layer B as shown by the arrows and further stretching along the the [1,0,0] direction from the distance $a$ to $a'$ by about 10-15\%.}
	\label{fig12}
\end{figure}
%\begin{figure}[!htb]
%\vskip-8mm
%	\includegraphics[width=1.1 \columnwidth]{CN_LI_HCP_ORT.pdf}
%	\vskip-8mm
%	\caption{(Color online) The mean number $n$ of close neighbors $n_n$ vs the radius $R_0$ of the coordination sphere in three samples: superfluid (green half filled hexagons); hcp solid (blue filled circles); orthorhombic solid (red stars).  The error bars represent r.m.s. of $n_n$. The horizontal lines and the numbers close to them indicate the flat part of the curves where $n=12, 14$ for the hcp and the orthorhombic samples, respectively. The numbers $n=18, 26$ mark the  second coordination sphere (which is much less pronounced). Inset: density of the superfluid calculated as $n/(4\pi R_0^3/3)$ vs $R_0$ (error bars due to the uncertainty in $n_n$ are not shown).  }
%	\label{fig2}
%\end{figure}

As mentioned above, the jog-antijog pair (the non-basal edge dislocations with their cores aligned with the hcp-axis) was observed to annihilate in long runs. Depending on the value of the chemical potential, this resulted in either gaining or loosing particles. In the first case,  the 1/2 atomic plane is restored only partially, and the resulting crystal structure is hcp slightly stretched along the X-axis. In the second case, the crystal structure changes radically. It is shown in Fig. \ref{fig11}. Its symmetry is face centered orthorhombic (fco), which is higher than that of the hcp stretched along the [1,0,0] direction. Such a crystal can be obtained from the hcp one by shifting the B layer with respect to the A layer by $a/2\sqrt{3}$ along the [0,1,0] direction, where $a$ is the hcp unit cell size along the [1,0,0] direction,  and further stretching along the same direction. [See the sketch in Fig.~\ref{fig12}]. The fco phase is likely induced by the uniaxial strain resulted from the annihilation of the jog-antijog pair. Further discussion of the structure as well as the analysis of the nature of the transition between the strained hcp and fco symmetries will be presented elsewhere.

\section{Discussion} 
The observed thermal activated response  is consistent with the general understanding that dislocations are surrounded by a cloud 
of point defects which contribute to dissipative motion of dislocations \cite{Hirth}. In most materials (at not too low $T$) the motion of vacancies can be described by the activated jumps. Accordingly, the dynamical response of the material is characterized by the energy which, in addition to the energy of creating a point defect, contains a part dealing with potential barriers for activated jumps. Thus knowing the activation energy determining the equilibrium concentration $n_p$ of the point defects is not sufficient for predicting the $T$-dependence of a dynamical response measured experimentally. The situation in solid \he4 is different. Vacancies in solid \he4 are quantum particles which, once created, can tunnel through the host lattice. At $T$ below the corresponding bandwidth, this tunneling can result in the diffusion coefficient $D$ which depends on $T$ as a power law (cf. Ref. \cite{Kagan_Maksimov}). Then, the dynamical response, which depends on the product $n_p D$, becomes controlled by the equilibrium activation energy $E_a$ in the main exponential approximation.   
Another aspect -- related to the grand canonical versus canonical ensembles -- must also be taken into account. In the GCE simulation, interstitials  and vacancies can be created independently from each other -- as Schottky-type defects. In a real experiment, the mass conservation must be respected. Then, in a perfect ideal uniform crystal, only Frenkel pairs can be created. This implies huge activation energies which essentially closes this channel for dynamics at temperatures below 2-3K. The situation is different in a real crystal characterized by spatial density variations induced by, e.g., a forest of dislocations, grains, container walls. A vacancy can be created in a region of low density and, then, tunnel to a region of higher density -- without creating an interstitial. Thus, the values of $E_a$ obtained here can be compared directly with the dynamical responses observed experimentally in realistic samples of solid \he4.

The activated mechanical response of solid \he4 is often related to pinning of the gliding dislocations by $^3$He impurities with typical activation energies below 1K:  0.7K (see in Ref.\cite{RMP}), 0.8K,Ref.\cite{Polturak_3_6} and varying from 0.18K to 0.35K, Ref.\cite{Iwasa}. In the solid \he4 of natural purity these impurities evaporate from the dislocations at temperatures above 0.2-0.3K, and the intrinsic mechanisms must take over.
The nature of such mechanisms is not well understood. In the ultrasound experiment \cite{Tsuruoka} the activation energies of 12-14K have been reported, and it has been suggested that these are related to phonon modes localized close to gliding (basal) dislocations. At this point we note that the reported values are close to the vacancy creation energy in the ideal \he4 obtained from the simulations  in Ref.\cite{fate} at the melting pressure. In this work we obtained e $E_a \approx 19$K in the ideal solid and $E_a \approx 15$K for basal dislocation (at the density 0.03$\AA^{-3}$, which is 2\% above the lowest density 0.0294 used in experiment \cite{Tsuruoka}). Thus, we can put forward an alternative mechanism for the activation response -- damping of the dislocation excitations by the cloud of the point defects bound to the dislocation (with the rate being proportional to the equilibrium density $n_p\sim \exp(-E_a/T)$ of the point defects).   

Another intrinsic defect which can contribute to the activation is a jog on a basal dislocation \cite{Hirth}. It plays a role of the dislocation pinning center and can be undone by absorption or emission of vacancies. The energy of the jog in solid \he4 has been estimated as $\approx 5.8$K in Ref.\cite{Tsuruoka}. However, no signature of this activation energy has been observed in Ref.\cite{Tsuruoka}. More recently, the activation energies of 5-6K have been reported in Refs.\cite{Beamish_5,Polturak_3_6}, and attributed to jogs of the basal dislocations in Ref.\cite{Beamish_5}. In this respect we note that $E_a =5.2 \pm 0.8$K,Fig.~\ref{fig10},  has been obtained in simulations of the non-basal edge dislocations with cores oriented along the Z-axis, and, thus, representing jogs of the basal dislocations. These values are consistent with each other.  

There is also a report of the activation energy $\approx 3$K in Ref.\cite{Polturak_3_6}. 
Among the defects simulated here there is one which is characterized by the activation energy close to this value. It is the partial dislocation of the E-fault with $E_a=4.1 \pm 0.3$K at $T\geq 1$K.   Such a dislocation, however, does not contribute directly to the plastic deformation along the basal plane. The question then is how can it contribute to the mechanical response?  As described in Ref. \cite{Kuklov2019}, a basal dislocation can bind to the dislocation with the Burgers vector along the hcp axis. This implies that  glide of the first can be strongly affected by the climb of the second, which is controlled by the point defects cloud formed around its core. It should also be taken into account that this dislocation exhibits two activation energies, with $E_a\approx 1$K for $T$ below 1.37K. That low values were not reported in the association with any intrinsic dissipation mechanism so far. However, this quantity should be compared to the 0.7K attributed to the binding energy of $^3$He impurities to basal dislocations (see in Ref.\cite{RMP}). To what extent this is just a coincidence remains to be seen -- especially within the view of the wide spread (by a factor of 2-4 (see in Ref.\cite{RMP}),\cite{Polturak_3_6,Iwasa}) of the reported $^3$He binding energies. 
It is also interesting to understand the nature of the recently reported activation energy 6.9 K \cite{Iwasa} (at the density $0.0297 \AA^{-3}$). It is not close to any value reported here for the density $0.03 \AA^{-3}$.

It is important to realize that, as $T$ lowers, the flutter induced friction \cite{flutter} dominates. The resulting friction has a contribution $\propto T^3$. Thus, there must be a temperature $T_f$ below which the cloud of the point defects bound to the dislocations becomes irrelevant for the dynamical response. In other words, the activation energy $\sim 1$K can well be obscured by the flutter mechanism. 

We should also mention that the dislocations analyzed here are far from exhausting all possible topological defects in solid \he4. The main result of this study is the demonstration of a wide variety of the intrinsic activation energies covering more than one order of magnitude -- from 1K to 20K depending on a dislocation type and the crystal density.   

Our simulations have revealed an unexpected feature of solid \he4 -- the orthorhombic phase which can be induced upon uniaxial deformation of about 10-15\%. This phase can be formed in local mesoscopic clusters due to the dislocation strain fields. This implies that, in general, when considering dislocation effects in hcp solid \he4, the structural transformation must be taken into account. The orthorhombic phase can also be detected in a controlled experiment applying uniaxial strain. This effect is an example of a martensitic transformation induced by the uniaxial stress similar to the recently observed orthorhombic to hexagonal close-packed martensite formation in high-entropy alloys \cite{mart}. Similar effect was observed in the Cu-Sn-Ga alloy \cite{UFN},
and it has been emphasized that such transitions are often characterized by, practically, undetectable or very low (as compared to melting) latent heat. This feature has never been reported in solid \he4 and deserves further numerical and experimental studies.

 {\bf Acknowledgment}. 
Authors acknowledge useful discussion of crystal structures with Alexander Zaitsev. This work was supported by the National Science Foundation under the grants DMR-1720251, DMR-2032136 and DMR-2032077.  Computations have been conducted at the ATLAS Israel Group supercomputing cluster. 
We acknowledge assistance by its system administrator David Cohen.

\end{document}